\documentclass[prb,twocolumn,aps,showpacs]{revtex4}
\usepackage{graphicx}
\usepackage{color,latexsym}
\def\be{\begin{equation}}
\def\ee{\end{equation}}

\begin{document}

\title{On the role of the symmetry parameter $\beta$ in the strongly localized regime}
\author{P. Marko\v{s}$^1$ and L. Schweitzer$^2$}
\affiliation{$^1$Department of Physics FEI, Slovak University of
  Technology, 812\,19 Bratislava, Slovakia\\
$^2$Physikalisch-Technische Bundesanstalt (PTB), Bundesallee
  100, 38116 Braunschweig, Germany}

\begin{abstract}
The generalization of the Dorokhov-Mello-Pereyra-Kumar equation for
the description of transport in strongly disordered systems replaces
the symmetry parameter $\beta$ by a new parameter $\gamma$, which
decreases to zero when the disorder strength increases. We show
numerically that although the value of $\gamma$ strongly influences
the statistical properties of transport parameters $\Delta$ and of the
energy level statistics, the form of their distributions always
depends on the symmetry parameter $\beta$ even in the limit of strong
disorder. In particular, the probability distribution is $p(\Delta)\sim
\Delta^\beta$ when $\Delta\to 0$ and $p(\Delta) \sim \exp (-c\Delta^2)$ in
the limit $\Delta\to\infty$. 
\end{abstract}


\maketitle

It has recently been shown by Muttalib {\itshape et al.}\cite{MG02,MK99} 
that the Dorokhov-Mello-Pereyra-Kumar Equation\cite{Dor82,MPK88}
(DMPKE) for the description of electronic transport in disordered
quasi-one-dimensional systems can be generalized to comprise also the
strongly disordered case. They found that the main difference
between the diffusive and localized regime is reflected in the spatial
distribution of the electrons inside the sample. The DMPKE was derived
under the assumption that the electron density is homogeneous. This
assumption is valid in the limit of weak disorder (diffusive regime)
and leads to universal behavior of the electron transport, which is
determined by only three parameters: the ratio $L_z/\ell$ of the system
length $L_z$ to the mean free path $\ell$, the number of scattering
channels $N$, 
and the symmetry parameter $\beta$ of the respective random matrix
ensemble. The latter determines the statistical properties of the the
model, for instance, the fluctuation of the conductance, var~$g\sim
\beta^{-1}$.   

The derivation of the DMPKE is based on the following representation 
of the transfer matrix,\cite{Mbook} which describes the scattering
of electrons coming from the left (right)
\be\label{f}
T = 
\left(
\begin{array}{ll}
u^{(1)} & 0\\
0     & u^{(2)}
\end{array}
\right)
\left(
\begin{array}{ll}
\sqrt{1+\lambda}  & \sqrt{\lambda}\\
\sqrt{\lambda}  &   \sqrt{1+\lambda}
\end{array}
\right)
\left(
\begin{array}{ll}
u^{(3)} & 0\\
0     & u^{(4)}
\end{array}
\right)
\ee
where $\lambda$ is a $N$-dimensional real diagonal matrix and the
structure of the matrices $u$ is given by the physical symmetry. 
The diagonal elements $\lambda_a$ ($a=1,2,\dots,N$) define the
conductance of the system via the Economou-Soukoulis formula\cite{ES81} 
\be\label{ES}
g = \frac{e^2}{h} \sum_a^N {\frac{1}{1+\lambda_a}}.
\ee
The probability distribution $p(\lambda)$ 
of parameters $\lambda_a$ is given by the DMPKE\cite{Dor82,MPK88}
\be\label{dmpk}
\frac{\partial {p}_{L_z}(\lambda)}{\partial (L_z/\ell)}
=\frac{2}{\beta N +2-\beta}
\frac{1}{{J}}\sum_a^N
\frac{\partial}{\partial\lambda_a}\left[\lambda_a(1+\lambda_a)
{J}\frac{\partial {p_{L_z}(\lambda)}}{\partial \lambda_a}\right]
\ee
with
${J}\equiv\prod_{a<b}^N|\lambda_a-\lambda_b|^{\beta}$.
The parameterization $\lambda_a=(\cosh x_a -1)/2$ introduces a new set
of variables $x_a$ ($x_1<x_2<\dots $), which follows the Wigner-Dyson
statistics. The probability distribution $p(\Delta_a)$ of the
normalized differences 
$\Delta_a = (x_{a+1}-x_a)/\langle x_{a+1}-x_a\rangle $ is well
described by the Wigner distribution\cite{Pic91} 
\be\label{wigners}
P_\beta(\zeta) = A_\beta \zeta ^\beta \exp (-B_\beta\zeta^2),
\ee
where for $\beta=1$, 2, 4: $A_1=\pi/2$, $B_1=\pi/4$,
$A_2=32/\pi^2$, $B_2=4/\pi$, $A_4=2^{18}/3^6\pi^3$, and $B_4=64/9\pi$,
respectively.

Several investigations showed\cite{AS86,AZKS88,Sea93} that the same
function also describes the probability distribution $p(s)$ of the
energy level statistics in the diffusive regime. 
Here, $s$ is the difference of consecutive energy eigenvalues
$s=|\varepsilon_{i+1}-\varepsilon_i|/\bar{s}$ divided by the mean level
spacing $\bar{s}$. 

In strongly disordered samples, the propagation of the electron is not
diffusive. We cannot expect that all paths across the sample are equivalent. 
Mathematically, this leads to the re-formulation of 
the DMPKE into the more general form   
\be\label{gdmpk}
\frac{\partial {p}_{L_z}(\lambda)}{\partial (L_z/\ell)}
=
\frac{1}{{J}}\sum_a^N
\frac{\partial}{\partial\lambda_a}\left[\lambda_a(1+\lambda_a)K_{aa}
{J}\frac{\partial {p_{L_z}(\lambda)}}{\partial \lambda_a}\right],
\ee
where
parameters $K_{ab}$ depend on the statistical properties of matrices $u$ 
in Eq.~(\ref{f}). 
The explicit form of $K_{ab}$ is determined by the model symmetry.
\cite{MMW,Mello,Chalker}  The Jacobian $J$ now have a form
\be
{J}\equiv\prod_{a<b}^N|\lambda_a-\lambda_b|^{\gamma_{ab}},
\quad \gamma_{ab}\equiv\frac{2K_{ab}}{K_{aa}}.
\ee
Although Eq.~(\ref{gdmpk}) was derived only for orthogonal systems
($\beta=1$), it can be shown to be valid also for $\beta=2$ and
$\beta=4$. The conductance is still given by Eq.~(\ref{ES}), it
becomes implicitly a function of the spatial distribution of the
electrons.  

The main difference between the DMPKE and its generalized version lies
in the presence of the parameters $\gamma_{ab}$ in the Jacobian. 
Later work showed\cite{MMW} that it is possible to approximate all
$\gamma_{ab}$ by a single parameter $\gamma$. Similarly, the
parameters $K_{aa}$ are substituted by a constant, which is of
order of unity in the limit of strong disorder. It was
argued\cite{MK99} that $\gamma\to \beta$ in the diffusive regime but
$\gamma\to 0$ when the disorder increases. This assumption 
was confirmed, at least for the orthogonal symmetry, by numerical
work.\cite{MMW}  

If $\gamma$ really decreased to zero, one would expect that the
probability distribution of $\Delta_a$ and that of the level
statistics should converge to the Poisson distribution 
\be
P_P(\zeta) = e^{-\zeta}.
\ee
Such changes of the distributions due to the increase of the disorder
are really observed both in the parameters $\Delta_a$\cite{MarkosKramer}
and in the level statistics. In the latter case it
has been used for the estimation of critical parameters of the
metal-insulator transitions.\cite{Sea93,SZ97,BSK98,PS02}

The role of $\gamma$ and its relation to the symmetry parameter
$\beta$ still requires a more detailed discussion. Therefore,  
we investigate in this paper the shape of numerically obtained
distributions $p(\zeta)$ in the limit of strong disorder. 
We show that, although $\gamma$ indeed decreases to zero,
both the distribution of $\Delta=(x_2-x_1)/\langle x_2-x_1\rangle$ 
and the level statistics $p(s)$ do depend on the symmetry parameter
$\beta$.  In particular, the small $\zeta$ behavior of these
distributions is always given by 
\be\label{small}
p(\zeta) \sim \zeta^\beta
\ee
independently on the strength of the disorder. However, this
power-law behavior is observed only in a very narrow range close to 
zero.

The systems to be investigated are defined on a 2D square lattice with
lattice constant $a_l$ and described by a tight-binding lattice
Hamiltonian with nearest-neighbor hopping terms 
\be\label{ham}
{\cal{H}}=\sum_{r,\sigma} \epsilon_{r\sigma} c_{r\sigma}^\dag c_{r\sigma}
+\sum_{\langle r\ne r'\rangle, \sigma\sigma'}
t_{rr'}^{\sigma\sigma'}c_{r\sigma}^\dag c_{r'\sigma'}. 
\ee
Here, $\sigma=\pm1/2$ is the electron spin, $r$ are the sites of the
2D lattice of size $L^2$, $\epsilon_r$ are the appertaining
random on-site energies distributed according to the box probability
$P_{\mathrm Box}(\epsilon_r)=(1/W)\Theta(W/2-|\epsilon_r|)$, and $W$
measures the strength of the disorder. Periodic boundary conditions
are applied in both directions. For the symplectic Ando model,
$t_{rr'}^{\sigma\sigma'}$ is a $2\times 2$ matrix with 
\begin{equation}
t_{xx'}^{\sigma\sigma'}=
\left(\matrix{
V_1\hfill &-V_2\hfill\cr
V_2\hfill & V_1\hfill\cr
}\right),~~~~
t_{yy'}^{\sigma\sigma'}=
\left(\matrix{
V_1\hfill &-iV_2\hfill\cr
-iV_2\hfill & V_1\hfill\cr
}\right)
\label{ando}
\end{equation}
and $V_1^2+V_2^2=V^2=1$. For the orthogonal model,
$t_{rr'} \equiv t_{rr'}^{\sigma\sigma'}\delta_{\sigma\sigma'}$. Energies
and lengths are measured in units of $V$ and lattice constant $a_l$,
respectively. For $E=0$ and $V_1=0.5$ the symplectic model exhibits a
metal-insulator transition at a critical disorder $W_c\approx
5.84$.\cite{MS06}   

The limiting behavior of the distribution $p(\zeta)$ is better visible
when the distribution of the logarithm of $\zeta$, $p(\ln\zeta)$, is
studied instead. From the equation
\be
p(\zeta)d\zeta = p(\xi)d\xi, \quad \xi=\ln\zeta
\ee
we obtain that the the relation $p(\zeta)\sim \zeta^\beta$ corresponds
to 
\be\label{origin}
\ln p(\xi) \sim (\beta+1)\xi, \quad \xi\to -\infty.
\ee
Similarly, the large-$\zeta$ tail of the distribution can be analyzed
from the function 
\be\label{tail}
\ln\left[-\ln p(\xi)\right] = \alpha\xi
\ee
with $\alpha=2$ and 1 for the Wigner and Poisson distribution, respectively.

\begin{figure}[t]
\includegraphics[clip,width=0.35\textwidth]{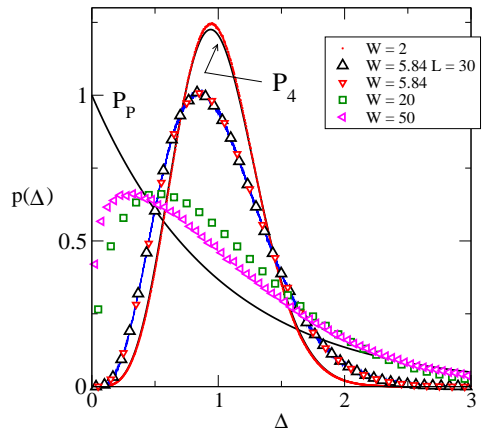}
\caption{(Color online) The distribution $p(\Delta)$ for the 2D Ando
  model in the metallic, critical ($W=5.84$), and localized
  regime. The linear size of the system is $L=14$. The data are
  compared with the Wigner surmise $P_{4}(\Delta)$ 
  and with the Poisson distribution $P_P(\Delta)$. While $p(\Delta)$ is
  very similar to the Wigner surmise in the metallic regime ($W=2$),
  it resembles the Poisson distribution when the disorder increases. 
  However, for any disorder strength, $p(\Delta)$ decreases
  to zero when $\Delta\to 0$. 
}
\label{AM_lin}
\end{figure}

In the limit of strong disorder, $x_1\gg 1$, the typical conductance
is given by the smallest parameter $x_1$ as $g \approx e^{-x_1}$. The
parameter $x_1$ determines the localization length $\lambda$ as
$x_1=2L/\lambda$ ($L\gg\lambda$). Thanks to this relation, the
transport properties of strongly disordered system can be understood
from the numerical analysis of relatively small samples, provided that
$L\gg\lambda$. 

\begin{figure}[t]
\includegraphics[clip,width=0.35\textwidth]{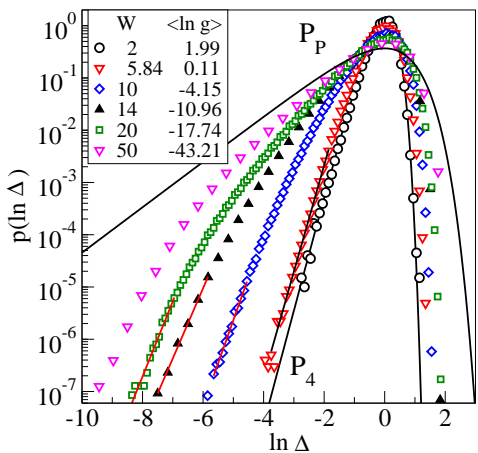}
\caption{(Color online) The distribution $p(\ln\Delta)$ for the 2D
  Ando model. The linear size of the samples is $L=14$. The solid
  curves represent the Wigner $P_{4}$ and the Poisson $P_P$
  distributions. With increasing disorder, the distribution of
  $p(\ln\Delta)$ changes. It is similar to Poissonian in the bulk, but
  for very small $\Delta$ we see the linear behavior $\ln p(\ln\Delta)
  \propto\ln\Delta$. The straight solid lines are fits for $W=5.84$,
  10, 14, and 20 with slopes 4.789, 3.78, 3.0, and 3.3, respectively.
}
\label{AM_log}
\end{figure}

\begin{figure}[b]
\includegraphics[clip,width=0.37\textwidth]{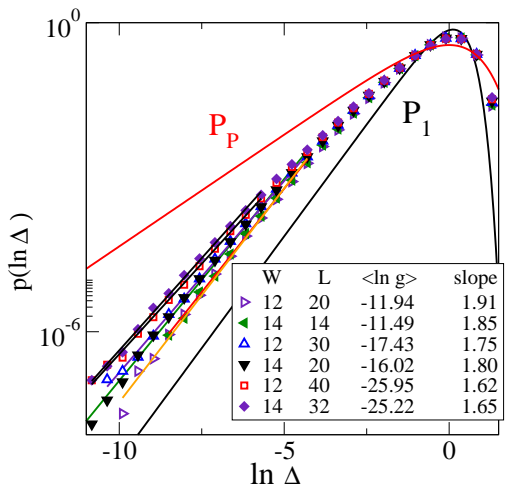}
\caption{(Color online) The distribution $p(\ln\Delta)$ for a 2D orthogonal
  model with anisotropic hopping ($t_{xx'}=0.9 t_{zz'}$). Two ensembles
  with different values of $L$ and $W$ but with the same value of   
  $\langle\ln g\rangle$ possess the same distribution $p(\Delta)$.
  The Wigner surmise $P_{1}$ and the Poisson
  distribution $P_P$ are also plotted. Fits shown by straight
  lines confirm that $\ln p(\ln\Delta)\propto \ln\Delta$
  with slopes given in the legend.
  Therefore, $p(\Delta)\sim \Delta$ when $\Delta\to 0$.
  The number of samples is $N_{\rm stat} = 1.6\times 10^9$ for
  $L=20$ ($W=12$), and about $10^8$ else.
}
\label{2D_log}
\end{figure}

\begin{figure}[t]
\includegraphics[clip,width=0.35\textwidth]{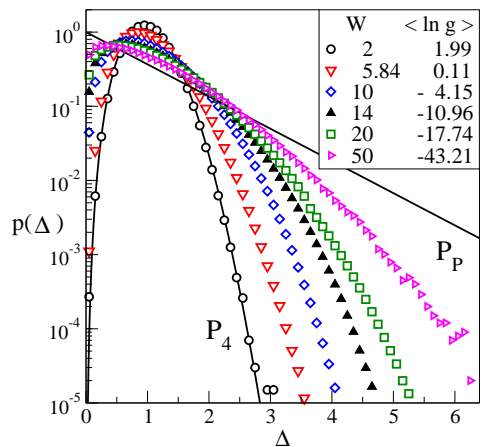}
\caption{(Color online) The same as in Fig.~\ref{AM_lin} but on a
  logarithmic scale for various strength of the disorder and
  $L=14$. Solid lines are Wigner and Poisson distributions. 
  For the localized regime ($W\ge 10$), the mean value of the logarithm
  of the conductance is given in the legend. 
}
\label{AM_lin1}
\end{figure}

We analyze statistical ensembles of $N_{\rm stat}\sim 10^8$ 
square samples\cite{FN}
(typical size is $L=14$) and collect the statistical 
distribution of the normalized difference. 
The results are displayed in Figs.~\ref{AM_lin}-\ref{AM_lin1} 
for the 2D Ando model and for the 2D orthogonal model. 
Fig.~\ref{AM_lin} exhibits the distribution $p(\Delta)$ for various
strengths of the disorder. The data show that for small disorder
($W=2$) the distribution is very similar to the Wigner surmise.
Although the form of the distribution changes when disorder increases,
the decrease $p(\Delta)\to 0$ is noticeable even for $W=50$. The
small-$\Delta$ behavior of the distribution is better visible in 
Fig.~\ref{AM_log} which plots the distribution $p(\ln\Delta)$. Our
numerical data for any disorder show that the distribution 
$p(\ln\Delta)$  becomes parallel to the Wigner surmise for very small
$\Delta$. This proves that $\ln p(\ln\Delta) \sim (\beta+1)\ln\Delta$ and,
consequently, $p(\Delta)\sim\Delta^\beta$ (Eq.~(\ref{origin})).
However, the power-like behavior $p(\Delta)\sim \Delta^\beta$ is
observed only for a very small part of the statistical ensemble. For
instance, the linear behavior $p(\ln\Delta)\sim (\beta+1)\ln\Delta$ is
observable only for $\ln\Delta < \ln\Delta_m = -4$ ($W=10$), $-7$
($W=20$) (Fig.~\ref{AM_log}). The  probability $p$ to find a system
with such a small value of $\Delta$ decreases rapidly when the disorder
increases:  $p\sim 10^{-3}$ ($W=10$) but $p\sim 10^{-6}$ for $W=20$.

\begin{figure}[b]
\includegraphics[width=0.38\textwidth]{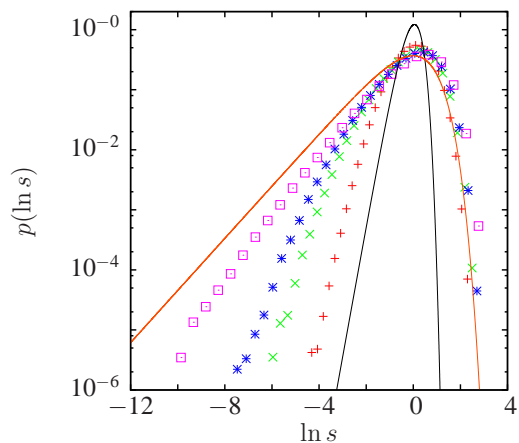}
\caption{(Color online) The energy level statistics $p(\ln s)$ of a strongly
  disordered Ando model. In the limit of $s\to 0$, a clear $\sim
  s^{\beta+1}$ behavior is observed for $W=10$ from the comparison
  with the Wigner surmise $P_{4}$ for symplectic symmetry. With
  increasing $W$ the range where this agreement holds shifts to
  smaller $s$. The Poisson fit (upper solid line) is valid only for
  larger $s$.  The system size is $L=20$ and the disorder strengths
  shown are $W=\mathrm{10}$ (+), 13 ($\times$), 15 ($\ast$), and 25
  ($\Box$),
  respectively. 
}
\label{pofls}
\end{figure}

According to single parameter scaling theory,\cite{AALR} the same
change of the distribution function is expected when the size $L$ of
the system increases while disorder is fixed.  
Owing to the necessity to analyze huge statistical ensembles, we did
not study the size dependence of $p(\Delta)$ for the Ando
model. However, we checked the $L$-dependence for 2D orthogonal systems
(Fig.~\ref{2D_log}).  Again, the distribution $p(\ln\Delta)$ follows
Eq.~(\ref{origin}) with $\beta = 1$ provided that $\Delta$ is
sufficiently small. Also, our data support the scaling idea: two
distributions are similar if they correspond to systems with the
same value of $\langle\ln g\rangle$. 

To estimate the large-$\Delta$ form of the distribution, we plot in
Fig.~\ref{AM_lin1}  $p(\Delta)$ for the Ando model and compare it with
the  Wigner distribution $P_{4}$ and the Poisson distribution. Again,
our data confirm that $p(\Delta)$ is never identical with the Poisson
distribution.  For large values of $\Delta$ the distribution is
$p(\Delta)\sim \exp (-c\Delta^b)$ with exponent $b\approx 2$, at least
in the limit of $\Delta\gg 1$.

The numerical investigation of the energy level statistics generated a
similar result. For large disorder, $W>W_c\simeq 5.84$, the large-$s$
part of the level statistics $p(\ln s)$ is well described by the Poisson
distribution as shown in Fig.~\ref{pofls}. In the opposite limit $s\to
0$, a behavior close to the Wigner surmise $P_{4}$ is observed in which
the range of the agreement is continuously diminishing with increasing
disorder. For very large $W>20$, however, only the downturn can still be
noticed. The eigenvalues have been calculated within an energy
interval $-0.5 \le \varepsilon_i \le 0.5$ by direct diagonalization of
the respective $(2L)^2 \times (2L)^2$ matrices with up to $3\times 10^5$
realizations. Additional calculations for larger system sizes
corroborated the results shown here for $L=20$.  

\begin{figure}[t]
\includegraphics[clip,width=0.36\textwidth]{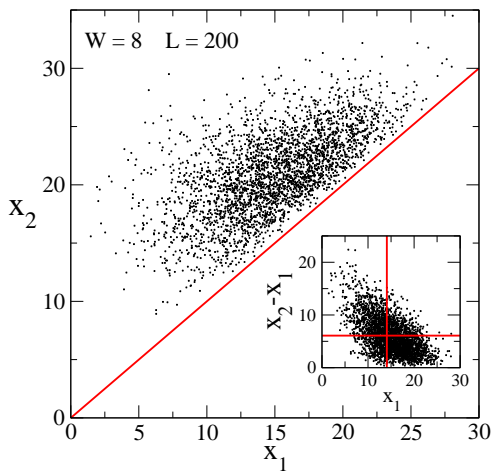}
\caption{(Color online) Plot of parameters $x_1$ and $x_2$ for a
  statistical ensemble of $10^4$ samples. The size is $L=200$ and
  disorder $W=8$. The inset shows the value of the difference
  $x_2-x_1$ as a function of $x_1$. Solid lines mark mean values,
  $\langle x_1\rangle=14.09$ and $\langle x_2-x_1\rangle=6.08$. 
The data confirm the level repulsion since $x_2$ and $x_1$
never coincide. Note that small values of $\Delta$ are observed only
in samples with relatively large value of $x_1>\langle x_1\rangle$. 
Such samples only marginally influence the mean conductance.
}
\label{AM_z2z1}
\end{figure}

Our numerical data indicate that the generalized DMPKE fails
to describe correctly the small-$\Delta$ behavior of $p(\Delta)$.
Contrary to the expected behavior $p(\Delta)\propto \Delta^\gamma$,
we observe for any disorder that $p(\Delta)\propto\Delta^\beta$.
This restriction influences the statistical properties of the
conductance only weakly since samples with small $\Delta$ represent  
only a very small part of the total statistical ensemble of the
$N_{\rm stat}$ samples. 
Also, as shown in Fig.~\ref{AM_z2z1}, samples with small $\Delta$ 
possess a large value of the smallest exponent $x_1> \langle
x_1\rangle$, and consequently have also a small conductance. 
Therefore, their contribution to the conductance statistics is
negligible. We expect that the GDMPKE gives a correct value of the 
mean $\ln g$ but its description of the small-$g$-tail of the
distribution $p(\ln g)$ can  eventually differ from  numerical
(experimental) data.

The aim of this paper was to verify that the physical symmetry of the
model governs the small $\zeta$ behavior of the distribution $p(\zeta)$,
where $\zeta$ denotes the normalized differences
$\Delta=x_2-x_1/\langle x_2-x_1\rangle$ or,
for the energy level statistics, the normalized difference of
consecutive eigenvalues $s=|\varepsilon_{i+1}-\varepsilon_i|/\bar{s}$.
This was done by a numerical study for the metallic and also in
the strongly localized regime. Our results confirm that the
generalized DMPK equation is not in contradiction with conclusions
provided by random matrix theory.  

We also showed that the distribution $p(\Delta)$ never corresponds entirely
to the Poisson distribution, although $\gamma\to 0$. $p(\Delta)$ is
not universal in the strongly disordered limit since $\gamma$ depends
on the disorder. For small $\Delta$ we found a distribution
$p(\Delta)\sim \Delta^\beta$ and in the limit of large $\Delta$ it
behaves as $\exp (-c\Delta^2)$. This observation is 
also consistent with the generalized DMPK equation. The Poisson
distribution indicates that the two parameters, $x_1$ and
$x_2$, are statistically independent. On the contrary, 
as discussed in previous work,\cite{M2000,MMW} the statistical
correlations survive for any disorder strength and are responsible
for the non-Gaussian distribution of the logarithm of the conductance.

PM thanks the Grant Agency VEGA for financial support of the Project
No.~0633/09.


\begin{thebibliography}{14}
\expandafter\ifx\csname natexlab\endcsname\relax\def\natexlab#1{#1}\fi
\expandafter\ifx\csname bibnamefont\endcsname\relax
  \def\bibnamefont#1{#1}\fi
\expandafter\ifx\csname bibfnamefont\endcsname\relax
  \def\bibfnamefont#1{#1}\fi
\expandafter\ifx\csname citenamefont\endcsname\relax
  \def\citenamefont#1{#1}\fi
\expandafter\ifx\csname url\endcsname\relax
  \def\url#1{\texttt{#1}}\fi
\expandafter\ifx\csname urlprefix\endcsname\relax\def\urlprefix{URL }\fi
\providecommand{\bibinfo}[2]{#2}
\providecommand{\eprint}[2][]{\url{#2}}

\bibitem[{\citenamefont{Muttalib and Klauder}(1999)}]{MK99}
\bibinfo{author}{\bibfnamefont{K.~A.} \bibnamefont{Muttalib}} \bibnamefont{and}
  \bibinfo{author}{\bibfnamefont{J.~R.} \bibnamefont{Klauder}},
  \bibinfo{journal}{Phys. Rev. Lett.} \textbf{\bibinfo{volume}{82}},
  \bibinfo{pages}{4272} (\bibinfo{year}{1999}).

\bibitem[{\citenamefont{Muttalib and Gopar}(2002)}]{MG02}
\bibinfo{author}{\bibfnamefont{K.~A.} \bibnamefont{Muttalib}} \bibnamefont{and}
  \bibinfo{author}{\bibfnamefont{V.~A.} \bibnamefont{Gopar}},
  \bibinfo{journal}{Phys. Rev. B} \textbf{\bibinfo{volume}{66}},
  \bibinfo{pages}{115318} (\bibinfo{year}{2002}).

\bibitem[{\citenamefont{Dorokhov}(1982)}]{Dor82}
\bibinfo{author}{\bibfnamefont{O.~N.} \bibnamefont{Dorokhov}},
  \bibinfo{journal}{Solid State Communications} \textbf{\bibinfo{volume}{41}},
  \bibinfo{pages}{431} (\bibinfo{year}{1982}).

\bibitem[{\citenamefont{Mello et~al.}(1988)\citenamefont{Mello, Pereyra, and
  Kumar}}]{MPK88}
\bibinfo{author}{\bibfnamefont{P.~A.} \bibnamefont{Mello}},
  \bibinfo{author}{\bibfnamefont{P.}~\bibnamefont{Pereyra}}, \bibnamefont{and}
  \bibinfo{author}{\bibfnamefont{N.}~\bibnamefont{Kumar}},
  \bibinfo{journal}{Ann. Phys. (N.Y.)} \textbf{\bibinfo{volume}{181}},
  \bibinfo{pages}{290} (\bibinfo{year}{1988}).

\bibitem{Mbook} P. A. Mello and N. Kumar, Quantum Transport in Mesoscopic Systems, Oxford Univ. Press
(2004), Oxford, UK

\bibitem[{\citenamefont{Economou and Soukoulis}(1981)}]{ES81}
\bibinfo{author}{\bibfnamefont{E.~N.} \bibnamefont{Economou}} \bibnamefont{and}
  \bibinfo{author}{\bibfnamefont{C.~M.} \bibnamefont{Soukoulis}},
  \bibinfo{journal}{Phys. Rev. Lett.} \textbf{\bibinfo{volume}{46}},
  \bibinfo{pages}{618} (\bibinfo{year}{1981}).

\bibitem[{\citenamefont{Pichard}(1991)}]{Pic91}
\bibinfo{author}{\bibfnamefont{J.-L.} \bibnamefont{Pichard}}, in
  \emph{\bibinfo{booktitle}{Quantum Coherence in Mesoscopic Systems}}, edited
  by \bibinfo{editor}{\bibfnamefont{B.}~\bibnamefont{Kramer}}
  (\bibinfo{publisher}{Plenum Press}, \bibinfo{address}{New York},
  \bibinfo{year}{1991}), vol. \bibinfo{volume}{254} of
  \emph{\bibinfo{series}{Nato ASI}}, pp. \bibinfo{pages}{369--399}.

\bibitem[{\citenamefont{Altshuler and Shklovskii}(1986)}]{AS86}
\bibinfo{author}{\bibfnamefont{B.~L.} \bibnamefont{Altshuler}}
  \bibnamefont{and} \bibinfo{author}{\bibfnamefont{B.~I.}
  \bibnamefont{Shklovskii}}, \bibinfo{journal}{Sov. Phys. JETP}
  \textbf{\bibinfo{volume}{64}}, \bibinfo{pages}{127} (\bibinfo{year}{1986}).

\bibitem[{\citenamefont{Altshuler et~al.}(1988)\citenamefont{Altshuler,
  Zharekeshev, Kotochigova, and Shklovskii}}]{AZKS88}
\bibinfo{author}{\bibfnamefont{B.~L.} \bibnamefont{Altshuler}},
  \bibinfo{author}{\bibfnamefont{I.~K.} \bibnamefont{Zharekeshev}},
  \bibinfo{author}{\bibfnamefont{S.~A.} \bibnamefont{Kotochigova}},
  \bibnamefont{and} \bibinfo{author}{\bibfnamefont{B.~I.}
  \bibnamefont{Shklovskii}}, \bibinfo{journal}{Sov. Phys. JETP}
  \textbf{\bibinfo{volume}{67}}, \bibinfo{pages}{625} (\bibinfo{year}{1988}).

\bibitem[{\citenamefont{Shklovskii et~al.}(1993)\citenamefont{Shklovskii,
  Shapiro, Sears, Lambrianides, and Shore}}]{Sea93}
\bibinfo{author}{\bibfnamefont{B.~I.} \bibnamefont{Shklovskii}},
  \bibinfo{author}{\bibfnamefont{B.}~\bibnamefont{Shapiro}},
  \bibinfo{author}{\bibfnamefont{B.~R.} \bibnamefont{Sears}},
  \bibinfo{author}{\bibfnamefont{P.}~\bibnamefont{Lambrianides}},
  \bibnamefont{and} \bibinfo{author}{\bibfnamefont{H.~B.} \bibnamefont{Shore}},
  \bibinfo{journal}{Phys. Rev. B} \textbf{\bibinfo{volume}{47}},
  \bibinfo{pages}{11487} (\bibinfo{year}{1993}).


\bibitem{Chalker} A. M. S. Macedo and J. T. Chalker, Phys. Rev. B 46, 14985 (1992)

\bibitem{MMW} K. A. Muttalib, P. Marko\v{s} and P. W\"olfle, Phys. Rev. B 72, 125317 (2005)

\bibitem{Mello} P. A. Mello, J. Phys. A 23, 4061 (1990)

\bibitem{MarkosKramer} P. Marko\v{s} and B. Kramer, Ann. Phys. 2, 339 (1993)

\bibitem[{\citenamefont{Schweitzer and {Kh. Zharekeshev}}(1997)}]{SZ97}
\bibinfo{author}{\bibfnamefont{L.}~\bibnamefont{Schweitzer}} \bibnamefont{and}
  \bibinfo{author}{\bibfnamefont{I.}~\bibnamefont{{Kh. Zharekeshev}}},
  \bibinfo{journal}{J. Phys.:\ Condens.\ Matter} \textbf{\bibinfo{volume}{9}},
  \bibinfo{pages}{L441} (\bibinfo{year}{1997}).

\bibitem[{\citenamefont{Batsch et~al.}(1998)\citenamefont{Batsch, Schweitzer,
  and Kramer}}]{BSK98}
\bibinfo{author}{\bibfnamefont{M.}~\bibnamefont{Batsch}},
  \bibinfo{author}{\bibfnamefont{L.}~\bibnamefont{Schweitzer}},
  \bibnamefont{and} \bibinfo{author}{\bibfnamefont{B.}~\bibnamefont{Kramer}},
  \bibinfo{journal}{Physica B} \textbf{\bibinfo{volume}{249-251}},
  \bibinfo{pages}{792} (\bibinfo{year}{1998}).

\bibitem[{\citenamefont{Potempa and Schweitzer}(2002)}]{PS02}
\bibinfo{author}{\bibfnamefont{H.}~\bibnamefont{Potempa}} \bibnamefont{and}
  \bibinfo{author}{\bibfnamefont{L.}~\bibnamefont{Schweitzer}},
  \bibinfo{journal}{Phys.\ Rev.\ B} \textbf{\bibinfo{volume}{65}},
  \bibinfo{pages}{201105(R)} (\bibinfo{year}{2002}).

\bibitem[{\citenamefont{Marko\v{s} and Schweitzer}(2006)}]{MS06}
\bibinfo{author}{\bibfnamefont{P.}~\bibnamefont{Marko\v{s}}} \bibnamefont{and}
  \bibinfo{author}{\bibfnamefont{L.}~\bibnamefont{Schweitzer}},
  \bibinfo{journal}{J. Phys. A} \textbf{\bibinfo{volume}{39}},
  \bibinfo{pages}{3221} (\bibinfo{year}{2006}).

\bibitem{FN} Although DMPKE was derived for
  quasi-one dimensional systems, there is strong numerical evidence
  that its predictions are applicable also for 2D and 3D
  systems\cite{MMW}

\bibitem{AALR} E.~Abrahams, P.~W.~Anderson, D.~C.~Licciardello, T.~V.~Ramakrishnan,
Phys.~Rev.~Lett. \textbf{42}, {673} (1979).

\bibitem{M2000} P. Marko\v{s}, Phys. Rev. B \textbf{65},  104207 (2002).

\end{thebibliography}

\end{document}